\documentclass{article}

\usepackage{arxiv}

\usepackage[utf8]{inputenc} 
\usepackage[T1]{fontenc}    
\usepackage{hyperref}       
\usepackage{url}            
\usepackage{booktabs}       
\usepackage{amsfonts}       
\usepackage{nicefrac}       
\usepackage{microtype}      
\usepackage{graphicx}
\usepackage{natbib}
\usepackage{doi}

\usepackage{mathpartir}
\usepackage{listings}
\usepackage{array,multirow,amsmath,amssymb,graphicx,array,url,subfig} 
\usepackage{xcolor}
\usepackage{float}
\usepackage{multirow}
\usepackage{adjustbox}

\title{A data-driven strategy to combine word embeddings in information retrieval}

\author{Alfredo Silva \\
	Department of Informatics \\
	Federico Santa Mar\'ia Technical University \\
	Santiago, Chile \\
	\texttt{alfredo.silva.13@sansano.usm.cl} \\
\And
\href{https://orcid.org/0000-0002-7969-6041}{\includegraphics[scale=0.06]{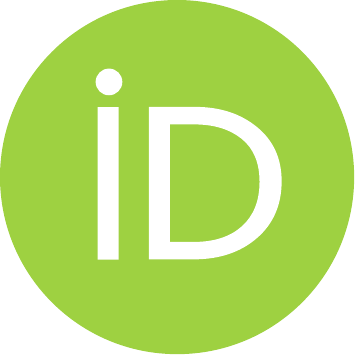}\hspace{1mm}Marcelo Mendoza} \\
	Department of Informatics \\
	Federico Santa Mar\'ia Technical University \\
	Santiago, Chile \\
	\texttt{marcelo.mendoza@usm.cl} \\
}

\renewcommand{\headeright}{Extended version}
\renewcommand{\undertitle}{Extended version}
\renewcommand{\shorttitle}{A data driven strategy to combine word embeddings}

\hypersetup{
pdftitle={A data driven strategy to combine word embeddings},
pdfauthor={Alfredo Silva, Marcelo Mendoza},
pdfkeywords={Word embeddings, Information retrieval},
}

\begin{document}
\maketitle

\begin{abstract}
	Word embeddings are vital descriptors of words in unigram representations of documents for many tasks in natural language processing and information retrieval. The representation of queries has been one of the most critical challenges in this area because it consists of a few terms and has little descriptive capacity. Strategies such as average word embeddings can enrich the queries' descriptive capacity since they favor the identification of related terms from the continuous vector representations that characterize these approaches. We propose a data-driven strategy to combine word embeddings. We use Idf combinations of embeddings to represent queries, showing that these representations outperform the average word embeddings recently proposed in the literature. Experimental results on benchmark data show that our proposal performs well, suggesting that data-driven combinations of word embeddings are a promising line of research in ad-hoc information retrieval.
\end{abstract}

\keywords{Word embeddings \and information retrieval \and query representation}

\section{Introduction}

Information retrieval is an essential area in information systems and document engineering. Its goal is to leverage query-document matching methods, from which a user can retrieve documents relevant to a query. These systems are organized into two components: a) Document retrieval, in which a set of documents potentially related to a query are retrieved, 2) Ranking: the retrieved documents are ordered using a ranking strategy. Usually, the document retrieval stage is performed by evaluating a bag-of-words query on an inverted index, from which the collection of documents containing the query words are retrieved. Then, the ranking phase is conducted using a word scoring function, which measures the relationship between the terms of the query and the terms of the document. Classic information retrieval schemes that follow this procedure are Tf-Idf \citep{SaltonB88} and BM25 \citep{RobertsonJ76}.

These types of methods have limitations. One of the most important is its inability to identify homonyms relationships between the words of the queries and the documents. Since the query-document matching scheme depends on a lexical match, it is not possible to retrieve documents that contain semantically related terms to the query if they do not have a lexical match. To address this limitation, many researchers have proposed automatic query expansion (AQE) strategies~\citep{CarpinetoR12,Azad:19}, adding related words to the query to retrieve more relevant documents.

AQE strategies define a candidate feature extraction method from a data source. Then by specifying a feature selection method, they select some potentially relevant terms to expand the query and reevaluate it. Usually, the data sources used for this purpose correspond to external sources, such as WordNet \citep{Voorhees94} or Wikipedia \citep{XuJW09}, or a local corpus built from the top-k ranked documents retrieved using the original query~\citep{DiazM06}. A feature selection method is used to define the level of relevance of each candidate word in the corpus with the query words. These methods cover a wide range of techniques and models to determine the significance of a term to the query. Among them, the literature reports methods based on mutual information \citep{HuDG06}, Kullback-Leibler divergence \citep{CarpinetoMRB01}, and probabilistic term-to-term associations \citep{CuiWNM03}, among others. Once the terms of the expansion are selected, the original query is combined with the new words to generate an expanded query. Different word-combining methods have been tried, such as boolean queries \citep{LiuLYM04}, unweighted term combination, or Rocchio term-weighting \cite{HeO07}. Once the expanded query representation is available, the expanded query is evaluated using a query-document matching scheme such as Tf-Idf \citep{SaltonB88} or BM25 \citep{RobertsonJ76}.

Polysemic words tackle techniques based on linguistic expansion that make use of external sources such as Wordnet. To correctly use a linguistic expansion technique based on, for example, homonyms, it is necessary to disambiguate the query according to its meaning. This task is difficult since the queries are short, and therefore very few information is available to do word sense disambiguation. The literature reports better results using all the terms of the query instead of expanding from single query words. This fact occurs because the use of the complete query provides more context to the expansion process. In general, better results are also shown using a local corpus for expansion, such as the top-k documents retrieved using the original query, instead of a global corpus such as Wikipedia. A local corpus shows advantages in this task since it focuses the technique on extracted features that are query-specific. Due to the above, query expansion techniques usually consider two stages of query evaluation, the first with the original query to determine the local corpus, and the second, in which the query is reevaluated using the terms of the expansion. Accordingly, these query-specific AQE methods need a double run at query time.

Representation learning has been a very active area in NLP in the last decade. Driven by a growing interest in the use of deep neural networks, several text representations have been proposed to solve different tasks supported on deep learning architectures. These representations operate at varying levels of aggregation, such as at the level of words, sentences, or documents. Word-level representations, known as word embeddings, allow building dense and low-dimensional vector representations. These embeddings allow identifying semantic relationships between words that do not have a lexical match. There are word embeddings like Word2Vec \citep{MikolovSCCD13} and GloVe \citep{PenningtonSM14} that provide one embedding for each word. This approach has limitations as it makes processing polysemic words difficult, and it is unable to construct representations for out-of-vocabulary words. The out-of-vocabulary problem has been tackled using subword-based methods such as FastText \citep{bojanowski2017enriching}, which can generate representations of out-of-vocabularies words from partial lexical matches. To address polysemy-based constraints, Elmo \citep{PetersNIGCLZ18} makes context-conditioned word embeddings. The use of the transformer architecture in BERT is of particular interest \citep{DevlinCLT19}. This architecture allows the detection of non-local dependency patterns in the text, being useful in different NLP tasks based on classification.

Word embeddings can be used to find term expansion candidates and then retrieving documents using the expanded query. Different strategies have been studied for estimating the relevance of candidate terms to the query comparing a single candidate term to every query term. The literature shows that word embedding based query expansion achieves competitive results when combined with strategies based on relevance feedback~\citep{ZamaniC16}. Like the classic AQE methods, the use of a local corpus based, for example, on the top-k documents retrieved from the original query, allows obtaining more relevant words for the meaning of the original query than if using a global vocabulary~\citep{Diaz:16}. For this reason, query-specific feature techniques predominate in AQE techniques based on word embeddings.

In this study, we compare different word embeddings strategies in AQE. We propose a data-driven strategy to combine word embeddings based on the IDF weights calculated on the corpus. Our approach aims to give more relevance to more informative terms, avoiding the inclusion of marginally relevant terms in the query expansion process.

The work is organized in the following way. In Section 2, we present related work. In Section 3, we discuss background knowledge on AQE. Our AQE strategy based on word embeddings is introduced in Section 4. Section 5 presents the experiments, the discussion of results, and the limitations of this work. Finally, we conclude in Section 6 providing concluding remarks and outlining future work.

\section{Related work}

\subsection{AQE strategies}

There are various approaches to designing AQE strategies. The first strategies that addressed this problem used linguistic resources, such as WordNet \citep{Miller92}, to identify the terms of the expanded query. These techniques looked for one-to-one relationships; that is, they found related terms by homonymy for each query term~\citep{Voorhees94}. WordNet organizes its words in synsets, so to choose the terms of the expansion, we must first select the synset most strongly related to the query term. This task requires a disambiguation process, called word sense disambiguation (WSD), which needs additional information to be addressed. Strategies based on associations generated automatically by computing term-to-term similarities in a collection of documents have been proposed~\citep{QiuF93}. These strategies address the problem of semantic ambiguity by retrieving the terms most strongly correlated with the query term in the corpus. For this purpose, term clustering and term to term correlation techniques have been evaluated~\citep{AlmasriBC13}. These approaches suffer from the word exchangeability problem, as they do not consider the order in which words appear in documents. Statistical strategies sensitive to the position of words in documents based on or word probabilities conditioned on observed terms have been studied in this problem \citep{BaiSBNC05}. Expansion features can also be generated by mining Wikipedia \citep{XuJW09} or user query logs, associating the terms of the original query with terms in past related queries~\citep{CuiWNM03,DupretM06}. All these techniques address the WSD problem assigning to the query the most frequent meaning in the collection.

Bai \textit{et al.}~\cite{BaiNCB07} shows that determining the meaning of the query is crucial in choosing the words for the expansion. For this reason, expansion strategies based on isolated terms tend to include marginally relevant terms. One way to address this limitation is to consider all the terms of the query to conduct the term search process~\citep{SunOC06}. In these types of search strategies, called one-to-many (one word relates to all the query terms), we might compute the correlation factors of a given candidate expansion term to every query term, and then combine the found scores to find the correlation to the global query. By combining multiple relationships between term pairs through Markov chains, Collins and Callan~\cite{Collins-ThompsonC05} builds a term network that contains pairs of words linked by several types of relations, such as synonyms, stems, co-occurrence, and transition probabilities. Then the words with the highest probabilities of relevance in the stationary distribution of the term network are selected as expansion features, reflecting the multiple aspects of the given query.

Another approach studied is to use an IR model to determine the top-k documents relevant to the query. Then, the top-k documents are used as a term search space. The idea of these strategies is to identify informative terms in the top-k documents, detecting differences between the term distribution of the (pseudo-) relevant documents, and the whole collection. Approaches based on mutual information~\citep{HuDG06}, Kullback-Leibler divergence~\citep{CarpinetoMRB01} or matrix factorization~\citep{Zamani:16} have shown interesting results in this research direction. An alternative to analysis based on distribution difference analysis is to build a query topic model from the top-ranked documents \citep{DiazM06}. The log-likelihood of top-ranked documents is computed combining the query topic model (to be estimated) and the collection language model. The approach assumes that query terms and documents are samples of the query topic model. The model can be estimated using the relevance model of Lavrenko and Croft~\cite{LavrenkoC01}. The approach makes documents that rank lower in the top-k list, have smaller influence in calculating word probabilities. WordNet has also been used to choose candidate expansion terms from a set of pseudo-relevant documents. Pal \textit{et al.}~\cite{Pal:14} determined that the overlap between the WordNet definitions of a candidate term and the query terms was useful in addressing the query expansion task. Probabilistic topic models were also used to choose the terms of an expansion in pseudo-relevance feedback scenarios. Colace \textit{et al.}~\cite{Colace:15} showed that topic models could help determine relevant word pairs, which were useful during the expansion process.

Once the terms of the expansion have been determined, AQE strategies use query re-weighting to generate a new representation of the query, including the terms of the expansion. This representation can be done using a term-scoring function, which combines the terms of the original query and the terms of the expansion in different proportions, controlled by a parameter. Typically, these strategies give higher weights to the terms of the original query, in a ratio of two to one~\citep{WongLLHL08}. In the case of AQE strategies based on query language models, the incorporation of the expansion terms is done directly in the relevance model~\citep{DiazM06}. More sophisticated techniques have also shown interesting results in query re-weighting tasks. Singh and Sharan~\cite{Singh:17} uses different term selection methods, combining them with varying weights using fuzzy rules to infer the importance of the additional query terms. This strategy has been especially useful for consolidating multiple term selection strategies in a single list of query expansion terms.

\subsection{AQE and word embeddings}

The use of neural word embeddings in information retrieval is recent. This line of research has had increasing interest from researchers and practitioners in the last five years. One of the first works in this area evaluated the effectiveness of skip-grams and CBOW in the translation language model \citep{Zuccon2015}. Both skip-grams and CBOW proved to be useful in ad-hoc information retrieval on TREC corpus. This success has prompted research on different IR tasks, such as sponsored search~\citep{ZhaiCZZ16} and web recommendations \citep{ManotumruksaMO18}.

The literature shows recent work in AQE based on word embeddings \citep{KuziSK16,RoyGBBM18,Diaz:16,ZamaniC16, ALMasri2016709}. Usually, these works use a pseudo-relevant feedback scenario (PRF), which is useful for restricting the term search space. Query expansion based on one-to-one relationships using word embeddings was studied by Almasri \textit{et al.}~\cite{ALMasri2016709}. The authors used cosine similarity to find terms related to each query term. Then a new query was built joining the identified words. Ad-hoc IR was evaluated using the query likelihood model on the expanded query. Results on CLEF data showed that their effectiveness was better than one-to-one AQE methods based on mutual information and PRF.

Other works build a representation of the query based on word embeddings. The terms for the expansion are identified using a similarity function between the vector representation of the query and the word embeddings of the terms that make up the (pseudo)-relevant documents. To construct the query vector representation, these methods average the word embeddings of the query terms, an approach known as average word embeddings (AWE). Both Roy \textit{et al.} \cite{RoyGBBM18} and Kuzi \textit{et al.} \cite{KuziSK16} studied the effectiveness of AWE vectors on AQE using cosine similarity, showing good performance when used in combination with PRF. Recently, Imani \textit{et al.} \cite{ImaniVMS19a} showed that cosine similarity can be replaced by a siamese network trained to detect candidate words.

Diaz \textit{et al.}~\cite{Diaz:16} incorporate relevance feedback in the process of learning the term embeddings. The idea is to retrieve a set of documents for the query to learn a query-specific term embedding model. Experimental results show that the terms of the query expansion identified using this procedure are more specific than those obtained by using embeddings trained on a global corpus. Along these same lines, Zamani and Croft~\cite{ZamaniC16} proposed a query likelihood model in which query embeddings are estimated in a local corpus. Using these data-driven query embeddings improves the effectiveness of the query expansion process. Driven by these promising results, Zamani and Croft~\cite{ZamaniC17} proposed relevance-based word embeddings, a word embedding strategy that learns word representations based on query-document relevance information. The authors showed that these word representations performed better on AQE than those based on word2vec or GloVe.

\section{Background}

The AQE strategy studied in this work is based on pseudo relevance feedback. Consequently, the AQE strategy considers two phases. In the first phase, the top-k documents are retrieved using an information retrieval model. In the second phase, we use the top-k documents retrieved as a local corpus. This corpus allows us to define the extraction phase of characteristics of the query in a query-specific context, focusing the AQE technique on a local vocabulary conditioned on the query. The assumption of relevance, or pseudo relevance feedback, is used by the AQE strategy to focus on the domain in which the candidate terms for expansion are sought.

Let $q$ be a bag-of-words query (BOW) composed by an arbitrary amount of terms that belong to a vocabulary $V$. Let $D = \{ d_1, \ldots , d_N \}$ be a collection of documents, or global corpus, in which the query $q$ is evaluated. A query engine retrieves the posting lists of the inverted index of $D$ for each word in $q$. After retrieving the posting lists, a set of documents $D^{'}$ is available. Then, for each document $d \in D^{'}$, a term-scoring function $\mbox{Term-score}(w_i, d)$ is used to rank the documents according to its relevance to $q$. Then, we calculate the ranking of $d$ by adding the term-scoring obtained for each word of $q$:

\[ \mbox{Score}(q,d) = \sum_{w_i \in q} \mbox{Term-score}(w_i, d). \]

Note that the term-scoring function may includes the IDF function. The literature shows many ways to calculate this function. We choose an expression that performs well in collections where the co-occurrence between terms and documents is sparse. Accordingly, IDF is given by the following expression:

\[ \mbox{Idf}(w_i) = \log \left( \frac{N - n(w_i) + 0.5}{N + 0.5} \right), \]

\noindent where $N$ is the number of documents in $D$ and $n(w_i)$ corresponds to the number of documents in which $w_i$ occurs. 

The $\mbox{Score}(q,d)$ function allows sorting the documents in $D^{'}$ according to their relevance to $q$. To build the local corpus on which the expansion of the $q$ query is made, the list of documents is truncated. Accordingly, the top-k documents in $D^{'}$ correspond to the local corpus from which the expansion terms will be searched. In this study, we work with a local corpus made up of the top-10 documents. Working with a small corpus allows us to focus the AQE strategy on a specific vocabulary, avoiding terms that are marginally relevant to $q$ can be considered in the expansion process. Using a small local corpus also avoids introducing high computational costs during the expansion process.

Our AQE strategy uses a query re-weighting model from which the terms of the query expansion can be identified. To do this, we build a query representation based on the word embeddings of the terms that compose it. This query representation is used to identify related terms. We use pre-trained word embeddings on large text collections to fulfill this purpose.

Many AQE strategies incorporate the terms of the expansion into a query likelihood model \citep{DiazM06,KuziSK16}. The idea of these models is to estimate the likelihood of a query term conditioned on a given document. In these models, the terms of the expanded query are incorporated into the likelihood calculation. Our expansion strategy is different since we use the terms of the expansion to re-evaluate the query in the BM25 model. We believe that the BM25 model can be useful to re-evaluate an expanded query in the document collection. The use of BM25 also facilitates the implementation of the AQE method in the query engine since it reuses the same query-document matching model used during the first phase of the AQE strategy.

\section{Proposal}

\subsection{IDF-AWE}

To identify the terms of the expansion, we created a representation of the query based on word embeddings. Both \cite{Diaz:16} and \cite{KuziSK16} use a similar approach named average word embeddings (AWE), which gives the same weight to each term of the query:

\begin{equation} 
\mbox{AWE}(q) = \frac{1}{n} \sum_{w_i \in q} \vec{w}_i,
\label{eq:0}
\end{equation}

\noindent where $\vec{w}_i$ is the word embedding of the query term $w_i$, and $n$ is the number of terms in $q$. Note that Equation \eqref{eq:0} can also be used to build representations of documents based on word embeddings.

Kuzi \textit{et al.}~\cite{KuziSK16} proposes the use of the closest term, measured by cosine similarity between the AWE vector and the word embedding of each candidate word, to determine which words will be incorporated in the query expansion. The authors define a scoring function for this purpose:

\begin{equation} 
\label{eq:1}
S(w,q) = \exp^{\cos(\vec{w}, \hspace{1mm} \mbox{AWE}(q))}, 
\end{equation}

\noindent where $w$ is a candidate word to be included in the query expansion process, and $\vec{w}$ is its word embedding. 

We extend AWE, introducing a data-driven strategy to combine the word embeddings. Our vector representation, IDF-AWE, is a representation of the query generated from the linear combination of word embeddings according to their IDF weights:

\begin{equation} 
\label{eq:2}
\mbox{IDF-AWE}(q) = \frac{1}{\sum_{w_i \in q}\mbox{IDF}(w_i)} \cdot \sum_{w_i \in q} \mbox{IDF}(w_i) \cdot \vec{w}_i, 
\end{equation}

\noindent where IDF is calculated with respect to the whole corpus. IDF-AWE will give more importance in representing $q$ to terms that are more specific in the corpus. To keep the magnitude of the word embeddings, the sum of the vectors is scaled by the sum of the factors IDF of the terms that belong to the query. In this way, the IDF-AWE vector gets the same magnitude as the word embeddings. Note that Equation \eqref{eq:2} can also be used to build representations of documents based on word embeddings.

Once the IDF-AWE vector is computed, we can use it to score the candidate terms to be used in the expansion. For this purpose, we use the score function defined in Equation \eqref{eq:1}. The expanded query will include the top-T terms closest to the IDF-AWE vector representation of $q$. The number of terms of the expansion, the value of T, corresponds to a parameter of the AQE strategy. These terms define a new query, denoted as $q_{exp}$, that will be re-evaluated in the query engine. 

\subsection{ElMo} The definition of the IDF-AWE vector for a context-dependent word embedding is different. In our study, we examined the effectiveness of ElMo \cite{PetersNIGCLZ18}, which defines a context-conditioned word embedding. To make use of this particularity that ElMo offers, we search for the occurrences of each query word in each document of $D^{'}$. Then, we use the surrounding text detected around each occurrence of $w_i$ as a context. Let $w_i$ be a word from the query $q$ and let $S_j(w_i, d) = \langle w_{i-N}, \ldots, w_i, \ldots w_{i+N} \rangle$ be a surrounding text of $w_i \in d$, whose length is $2N + 1$. By providing ElMo with the context $S_j(w_i, d)$, we obtain a context-conditioned word embedding, which we will denote $\vec{w}_{i,j}$. Let's assume that $w_i$ has $M$ matches in $d$ and therefore, ElMo returns $M$ word embeddings $\vec{w}_{i,j}, j \in \{1, M\}$. We obtain a word embedding for $w_i$ conditioned on $d$ by averaging the ElMo word embeddings obtained in $d$:

\[ \vec{w}_i = \frac{1}{M} \cdot \sum_{j=1}^M \vec{w}_{i,j}. \]

Once the ElMo word embedding for $w_i$ conditioned on $d$ is obtained, we can compute the IDF-AWE vector using the Equation \eqref{eq:2}. The length of the context window was set to 5 $(N=2)$, as usual in word embeddings~\citep{MikolovSCCD13}.

The retrieval phase of relevant documents from the expanded query $q_{exp}$ is implemented using the disjunction of the terms included in the expansion. This variant that incorporates our AQE strategy retrieves the posting list of the inverted index of $D$ for each word of $q_{exp}$. Instead of intersecting the postings lists, it retrieves all the documents addressed in these lists. Since the IDF-AWE vector gives more relevance to the more specific terms in $D$, the words of the expansion should be strongly related to these specific words. We hypothesize that by giving more relevance to specific words in the representation of $q$, the words in $q_{exp}$ will also be more specific. If this is effective, the AQE strategy is expected to improve the effectiveness of the system by joining the posting lists of these terms.

The ranking function defined in Equation 3 provides a high ranking to documents that are relevant to both query versions. However, Equation 3 can also ranks high  documents that rank high for $q$ or $q_{\mbox{new}}$. The relative importance that our AQE strategy gives to each query is defined by $\alpha$, which defines in which proportion both ranking functions are combined in the global ranking.

\subsection{Re-ranking}

Let $\mbox{top-T}(q) = \{ t_1, \ldots, t_{T} \}$ be the terms of the expansion of $q$. Let $p(t_i)$ be the posting list of each term $t_i$ in $\mbox{top-T}(q)$. A new set of documents is defined by  $D^{\mbox{exp}} = \{ \cup p(t_i) | \forall t_i \in \mbox{top-T}(q) \}$, that is, joining the posting lists of the terms of the expansion. Our AQE strategy ranks $D^{\mbox{exp}}$, applying a term-scoring function to each term of $q_{exp}$ and each document in $D^{\mbox{exp}}$. Once the documents in $D^{\mbox{exp}}$ are ranked, a global ranking of the documents in $D^{'} \cup D^{\mbox{exp}}$ is obtained, applying a re-ranking function. Our re-ranking function combines the rankings of both ordered lists of relevant documents, using a linear combination of the rankings obtained in each of the stages of the AQE strategy. 

Let $d$ be a document in $D^{'} \cup D^{\mbox{exp}}$. The scoring function for $d$ in our AQE system is given by:

\begin{equation}
    \label{eq:3}
    \mbox{Score}(\langle q , q_{exp} \rangle,d) = (1 - \alpha) \cdot \mbox{Score}(q,d) + \alpha \cdot \mbox{Score}(q_{exp},d), 
\end{equation}

\noindent where $\alpha \in [0,1]$ controls the relative importance of $q$ and $q_{exp}$ in the global ranking. Note that $\mbox{Score}(q,d) = 0$ if $d \notin D^{'}$ and $\mbox{Score}(q_{exp},d) = 0$ if $d \notin D^{\mbox{exp}}$. 

We introduce a variant of this re-ranking function, which, instead of combining the document-level rankings, modifies the BM25 ranking function according to the linear combination factors. To do this, each document is evaluated in a new query that has the terms of the original query and those of the expanded query. If the query term belongs to the original query, its BM25 term-scoring is weighted according to $(1- \alpha)$. If the query term belongs to the expanded query, its BM25 term scoring is weighted according to the $\alpha$ value. The following expression gives the re-ranking function:

\begin{eqnarray}
\label{eq:4}\nonumber
    \mbox{Score}(q \cup q_{exp},d) &=& (1 - \alpha) \cdot \sum_{w_i \in q} \mbox{BM25}(w_i,d) \\ 
    &+& \alpha \cdot \sum_{w_i \in q_{exp}} \mbox{BM25}(w_i,d).
\end{eqnarray}

\section{Experiments}

\subsection{Datasets}

The data available to evaluate the proposed strategies correspond to two partitions of the Tipster corpus, also known as the Text Research Collection Volume (TREC). The two Tipster partitions studied in this section are: Associated Press (AP) and Wall Street Journal (WSJ). Table \ref{tab:1} shows some basic statistics of these collections.

\begin{table}[h!]
    \caption{Collections used in this study.}
    \centering
    \resizebox{0.65\columnwidth}{!}{
    \begin{tabular}{|l|l|l|l|l|} \hline
    Corpus & Years & Documents & Tokens  & Vocabulary \\ \hline
    WSJ & 1987 - 1992 & 173252 & 81 $\cdot 10^6$ & 410578 \\
    AP & 1988 - 1990 & 239302 & 114 $\cdot 10^6$ & 417490 \\ \hline
    \end{tabular}}
    \label{tab:1}
\end{table}

The queries used in this study correspond to the enumerated TREC queries 51-200, of which there are also qrels that allow evaluating the effectiveness of the proposed strategies. The 150 test queries were evaluated in the three datasets for each of the experimental configurations studied in this section. All corpora in Table \ref{tab:1} were stopped using
the SMART stopword list and stemmed using the Krovetz algorithm.

\subsection{Word embeddings used in this study}

We study five strategies of word embeddings. Two variants of word2vec~\cite{MikolovSCCD13}, Skip-grams, and CBOW, were studied in this work. Both strategies were applied to a corpus that joins the two TREC datasets evaluated (WSJ + AP). Both CBOW and Skip-grams were trained using sliding windows with five terms, using an FFNN with a hidden layer of 300 units.

For GloVe, FastText, and ElMo, we use pre-trained vectors. GloVe vectors \cite{PenningtonSM14} were pre-trained on Wikipedia 2014 + Gigaword 5, two large-scale text collections. GloVe vectors also have 300 dimensions and it has a vocabulary of 400,000 terms. GloVe's match on each dataset is 89,142 terms in SJMN, 103,700 terms in WSJ, and 120,218 in AP.

FastText vectors \citep{bojanowski2017enriching} were pre-trained in a corpus of over 2.5 million materials science articles. FastText vectors have 100 dimensions. Since they are generated from subwords, they produce a full match with the TREC dataset vocabularies.

ElMo vectors \citep{PetersNIGCLZ18} were pre-trained in a corpus called 1 Billion Word Language Model Benchmark. ElMo vectors have 1024 dimensions. These vectors also generate a full match with the TREC dataset vocabularies.

\subsection{Experimental setting}

We study different variants of query-document matching methods that can be implemented using the concepts discussed in Sections 3 and 4. These variants allow us to identify the impact of each element of the proposal on the effectiveness of the AQE strategy. To illustrate the usefulness of each building block of the AQE strategy, we defined five variants, two of which do not implement AQE, and three of which implement different variants of the AQE strategy. Each of these variants is detailed below:

\vskip 1mm

\noindent AWE-VS: In this variant of the proposal, called AWE-based vector space, the documents in $D^{'}$ are ranked using the cosine similarity between the AWE vector of the query and the AWE vector of each document in $D^{'}$. AWE-VS does not consider query expansion. AWE-VS is evaluated using four word embedding strategies; these are, Skip-grams, CBOW, GloVe, and FastText. ElMo is not evaluated in this variant of the model since the construction of the vector representation of documents based on context-dependent encoding is very costly in computational time.

\vskip 1mm
 
\noindent IDF-AWE-VS: In this variant of the proposal, called IDF-AWE-based vector space, the documents in $D^{'}$ are ranked using the cosine similarity between the IDF-AWE vector of the query and the IDF-AWE vector of each document of $D^{'}$. IDF-AWE-VS does not consider query expansion. It is evaluated in CBOW, Skip-grams, GloVe, FastText, and ElMo. In the case of ELMo, it is more difficult to obtain a vector representation of the document since ElMo is context-dependent. In this configuration, all the occurrences on the document of each query term were searched. The AWE vector of the document context was obtained from them, using a term window of 5 words around the query term. Then, the context AWE vectors for each query term were averaged to generate one AWE vector per query term. The IDF-AWE vector of the document was obtained by combining the AWE vectors per word according to their Idf weights.

\vskip 1mm

\noindent AQE-Cent: AQE-Cent uses the AWE vector of $q$ to identify the terms of the expansion using the term-scoring function defined in Equation \eqref{eq:1}. Five variants of word embeddings are evaluated to construct the representation of the query (CBOW, Skip-grams, GloVe, FastText, and ElMo). The variants based on CBOW and Skip-grams correspond exactly to those studied by Kuzi \textit{et al.}~\cite{KuziSK16}. The candidate terms of the expansion correspond to the terms of the top-10 documents of $D^{'}$ according to the BM25 score obtained for the original query. The term-scoring function allows defining T terms related to $q$, which make up a new query. This new query is evaluated in $D$, and its results are ranked according to BM25. The results of both rankings were consolidated in a global ranking using the re-ranking function defined in Equation \eqref{eq:3}. 

\vskip 1mm

\noindent IDF-AWE-VS + AQE-Cent: This variant also uses AQE-Cent to represent the query during the expansion process. However, instead of searching for the terms in the top-10 ranked documents of $D^{'}$ according to BM25, it ranks them using IDF-AWE-VS. In this way, the search space for terms is different, giving more prominence to documents that are close to the IDF-AWE query vector. IDF-AWE-VS + AQE-Cent is evaluated using Skip-grams, CBOW, GloVe, and FastText. The terms of the expansion are determined using the term-scoring function defined in Equation \eqref{eq:1}. Then, the expanded query is evaluated in $D$, and its documents are ranked using BM25. The results of both rankings are consolidated in a global ranking using the re-ranking function defined in Equation \eqref{eq:3}.

\vskip 1mm

\noindent IDF-AWE-VS + AQE-IDF-Cent: This variant uses IDF-AWE-VS to rank the documents in $D^{'}$. The query representation is built using IDF-AWE, and the terms of the expansion are determined using the term-scoring function defined in Equation 1. Five variants of word embeddings are evaluated to construct the representation of the query (CBOW, Skip-grams, GloVe, FastText, and ElMo). The expanded query is evaluated in $D$, and its documents are ranked using IDF-AWE. The results of both rankings are consolidated in a global ranking using the re-ranking function defined in Equation \eqref{eq:3}. Another variant of this strategy that we study is to use the re-ranking function defined in Equation \eqref{eq:4}. This variant is indicated as IDF-AWE-VS + AQE-IDF-Cent$^+$.

The AQE strategies were studied using expansions with five terms. Top-k document retrieval on $D^{'}$ was conducted using the top-10 highly ranked documents in each experimental setting. We tested many values for $\alpha$, but $\alpha$ at 0.3 was the one with the best results. We included two baselines to carry out the evaluations. These are BM25 for the methods without expansion and BM25-AQE for the methods with query expansion. BM25-AQE uses the cosine similarity of each query term and each word embedding in the top-k documents to determine the terms of the expanded query. Once the expanded query has been evaluated, its results are ordered using BM25. We compare these results with a state-of-the-art method from the literature, the query-likelihood model with query expansion (QLM) introduced by Diaz \textit{et al.}~\cite{Diaz:16} and discussed in Section 3. 

The methods were evaluated using Mean Average Precision (MAP@10), Recall (R@10), and NDCG (NDCG@10). The evaluation @10 is usual in the validation of AQE strategies as it is expected that the effectiveness of the technique will be shown in the top-k documents of the ranking. The reported results correspond to averages across queries, for each corpus.

\subsection{Results}

The experimental results for WSJ and AP are shown in Table \ref{tab:2}. MAP and recall are shown in percentages and NDCG in the interval [0,1].

\begin{table}[h]
\caption{Experimental results in WSJ and AP.}
    \centering
    \resizebox{\columnwidth}{!}{
    \begin{tabular}{|l|c|c|c|c|c|c|} \hline
    & \multicolumn{3}{c|}{WSJ}
 & \multicolumn{3}{c|}{AP} \\ \hline
 \footnotesize{Method}       	&	 \footnotesize{MAP@10} 	&	 \footnotesize{R@10} 	&	 \footnotesize{NDCG@10} 	&	 \footnotesize{MAP@10} 	&	 \footnotesize{R@10} 	&	 \footnotesize{NDCG@10} \\ \hline
\footnotesize{BM25 \citep{RobertsonJ76}} 	&	 21,79 	&	 1,58 	&	0,54	&	 22,67 	&	 1,50 	&	 0,59 \\
\footnotesize{AWE-VS + CBOW } 	&	 25,59 	&	 1,85 	&	0,65	&	 23,48 	&	 1,68 	&	 0,60 \\
\footnotesize{AWE-VS + SG } 	&	 25,87 	&	 2,15 	&	 \textbf{0,70} 	&	 23,73 	&	 1,90 	&	 0,62 \\
\footnotesize{AWE-VS + GloVe } 	&	 21,93 	&	 1,46 	&	0,54	&	 23,31 	&	 1,27 	&	 0,50 \\
\footnotesize{AWE-VS + FastText } 	&	 21,77 	&	 1,20 	&	0,52	&	 23,26 	&	 1,29 	&	 0,51 \\
\footnotesize{IDF-AWE-VS + CBOW } 	&	 27,70 	&	 2,04 	&	0,67	&	 22,16 	&	 1,79 	&	 0,62 \\
\footnotesize{IDF-AWE-VS + SG } 	&	 27,09 	&	 2,16 	&	0,68	&	 23,37 	&	 \textbf{1,93} 	&	 \textbf{0,63} \\
\footnotesize{IDF-AWE-VS + GloVe } 	&	 23,65 	&	 1,44 	&	0,54	&	 23,54 	&	 1,33 	&	 0,49 \\
\footnotesize{IDF-AWE-VS + FastText } 	&	 21,31 	&	 1,34 	&	0,54	&	 20,88 	&	 1,32 	&	 0,52 \\
\footnotesize{IDF-AWE-VS + ElMo } 	&	 38,96 	&	 1,82 	&	0,46	&	 37,95 	&	 1,56 	&	 0,43 \\
\footnotesize{BM25 + AQE } 	&	 21,79 	&	 1,58 	&	0,54	&	 22,67 	&	 1,50 	&	 0,59 \\
\footnotesize{QLM \citep{Diaz:16}} 	&	 38,46 	&	 1,86 	&	0,57	&	 34,32  	&	 1.86 	&	 0,59 \\
\footnotesize{AQE-Cent + CBOW \citep{KuziSK16}} 	&	 24,42 	&	 1,51 	&	0,55	&	 26,68 	&	 1,55 	&	 0,61 \\
\footnotesize{AQE-Cent + SG \citep{KuziSK16}} 	&	 22,52 	&	 1,58 	&	0,59	&	 24,18 	&	 1,66 	&	 0,61 \\
\footnotesize{AQE-Cent + GloVe } 	&	 21,07 	&	 1,50 	&	0,59	&	 22,73 	&	 1,44 	&	 0,58 \\
\footnotesize{AQE-Cent + FastText } 	&	 20,97 	&	 1,44 	&	0,55	&	 22,19 	&	 1,54 	&	 0,58 \\
\footnotesize{AQE-Cent + ElMo } 	&	 24,42 	&	 1,51 	&	0,55	&	 23,17 	&	 1,63 	&	 0,60 \\
\footnotesize{IDF-AWE-VS + AQE-Cent + CBOW } 	&	 49,35 	&	 1,67 	&	0,49	&	 49,59 	&	 1,65 	&	 0,43 \\
\footnotesize{IDF-AWE-VS + AQE-Cent + SG } 	&	 58,25 	&	 1,86 	&	0,5	&	 59,72 	&	 1,78 	&	 0,43 \\
\footnotesize{IDF-AWE-VS + AQE-Cent + GloVe } 	&	 39,30 	&	 1,03 	&	0,29	&	 62,93 	&	 1,01 	&	 0,26 \\
\footnotesize{IDF-AWE-VS + AQE-Cent + FastText } 	&	 23,58 	&	 0,91 	&	0,41	&	 26,21 	&	 0,63 	&	 0,29 \\
\footnotesize{IDF-AWE-VS + AQE-IDF-Cent + CBOW } 	&	 48,71 	&	 1,20 	&	0,38	&	 54,44 	&	 1,15 	&	 0,36 \\
\footnotesize{IDF-AWE-VS + AQE-IDF-Cent + SG } 	&	 48,87 	&	 1,22 	&	0,38	&	 55,38 	&	 1,13 	&	 0,37 \\
\footnotesize{IDF-AWE-VS + AQE-IDF-Cent + GloVe } 	&	 48,09 	&	 1,19 	&	0,38	&	 54,33 	&	 1,12 	&	 0,37 \\
\footnotesize{IDF-AWE-VS + AQE-IDF-Cent + FastText } 	&	 49,30 	&	 1,22 	&	0,38	&	 58,65 	&	 1,14 	&	 0,37 \\
\footnotesize{IDF-AWE-VS + AQE-IDF-Cent$^++$ CBOW } 	&	 63,77 	&	 2,01 	&	0,46	&	 67,28 	&	 1,75 	&	 0,39 \\
\footnotesize{IDF-AWE-VS + AQE-IDF-Cent$^++$ SG } 	&	 64,01 	&	 2,00 	&	0,46	&	 66,79 	&	 1,76 	&	 0,40 \\
\footnotesize{IDF-AWE-VS + AQE-IDF-Cent$^++$ GloVe } 	&	 64,90 	&	 2,09 	&	0,46	&	 68,46 	&	 1,73 	&	 0,39 \\
\footnotesize{IDF-AWE-VS + AQE-IDF-Cent$^++$ FastText } 	&	 \textbf{68,10} 	&	 \textbf{2,19} 	&	0,48	&	 \textbf{69,92} 	&	 1,78 	&	 0,38 \\
\footnotesize{IDF-AWE-VS + AQE-IDF-Cent$^++$ ElMo } 	&	 64,07 	&	 2,09 	&	0,45	&	 66,40 	&	 1,76 	&	 0,38 \\ \hline
    \end{tabular}}
    \label{tab:2}
\end{table}

The results in Table \ref{tab:2} show that the strategies that use IDF-AWE-VS to determine the term search space consistently perform well. By combining this technique with the IDF-AWE query representation, the results are improved in terms of MAP. The best result obtained by this combination, indicated as IDF-AWE-VS + AQE-IDF-Cent, is obtained using FastText. This result is consistent in the three corpus studied. The MAP of IDF-AWE-VS + AQE-IDF-Cent$^+$ is the best observed in each corpus, surpassing all its competitors. The margin with which IDF-AWE-VS + AQE-IDF-Cent$^+$ outperforms AQE-Cent is remarkable, exceeding 30 percentage points of MAP in various configurations. This finding shows that the IDF-AWE query vector is very useful in query expansion tasks.

\begin{table}[h!]
\scriptsize
    \centering
    \caption{Query expansion terms detected using Skip-grams and FastText.}
    \resizebox{\columnwidth}{!}{
    \begin{adjustbox}{angle=90}
    \begin{tabular}{|l|l|l|l|} \hline
\multicolumn{4}{|c|}{Skip-grams} \\ \hline   
qid	&	query	&	AWE (AQE-Cent)	&	IDF-AWE	(IDF-AWE-VS + AQE-IDF-Cent$^+$) \\ \hline
101	&	design, star, war, anti-missil, defens, system	&	space-bas, sdi, antimissil, spacebas, effort-system	&	space-bas, sdi, \colorbox{cyan}{missile-defens}, antimissil, earlier-than-contempl	\\
102	&	laser, research, applic, u.s., strateg, defens	&	spaced-bas, early-deploy, 200-person-strong, defense-research, far-term	&	spaced-bas, early-deploy, directed-energi, defense-research, \colorbox{cyan}{sdi}	\\
103	&	welfar, reform	&	alongget, job-support, social, income-mainten, educ	&	alongget, social, job-support, income-mainten, educ	\\
104	&	catastroph, health, insur	&	higher-priv, health-car, policyhold, care, health-insur	&	\colorbox{cyan}{medicar}, higher-priv, \colorbox{cyan}{catastrophic-il}, health-car, catastrophic-car	\\
105	&	black, monday	&	tuesday, wednesday, thursday, friday, today	&	tuesday, friday, today, white, \colorbox{cyan}{hispan}	\\
106	&	u.s., control, insid, trade	&	also, u., public-investor, howev, although	&	\colorbox{cyan}{outsid}, also, although, howev, moreov	\\
107	&	japanes, regul, insid, trade	&	public-investor, mega-play, also, howev, seijirou	&	public-investor, \colorbox{cyan}{semi-insid}, mega-play, seijirou, \colorbox{cyan}{outsid}	\\
108	&	japanes, protectionist, measur	&	protection, brazilian-u.s., japan, nontariff, gatt-ord	&	protection, brazilian-u.s., \colorbox{cyan}{free-trad}, japan, \colorbox{cyan}{super-301}	\\
109	&	find, innov, compani	&	inde, corporate-awar, howev, instanc, also	&	idea, inde, \colorbox{cyan}{success}, corporate-awar, creativ	\\
110	&	black, resist, against, south, african, govern	&	political-reconstruct, africa, taxi-own, udf-link, anc-link	&	africa, political-reconstruct, taxi-own, udf-link, \colorbox{cyan}{pretoria}	\\
111	&	nuclear, prolifer	&	weapon, weapons-cap, warhead, weapon-fre, reactor	&	weapon, weapons-cap, \colorbox{cyan}{nuclear-weapon}, weapon-fre, warhead	\\
112	&	fund, biotechnolog	&	single-sector, +107.21, technolog, invest, 10th-best-perform	&	technolog, amgen, \colorbox{cyan}{pharmaceut}, sunagra, \colorbox{cyan}{repligen}	\\
113	&	new, space, satellit, applic	&	44p, intelsat-5, 111c, reboost, orbit	&	44p, orbit, payload, intelsat-5, 111c	\\
114	&	non-commerci, satellit, launch	&	44p, intelsat-5, earth-map, 4,580-pound, ro1	&	intelsat-5, 44p, \colorbox{cyan}{military-spac}, direct-broadcast, \colorbox{cyan}{military-satellit}	\\
115	&	impact, 1986, immigr, law	&	farm-wag, 377-a-year, joint-petit, immigration-reform, cuban/haitian	&	377-a-year, farm-wag, joint-petit, immigration-reform, \colorbox{cyan}{alien}	\\
116	&	generic, drug, substitut	&	macrocryst, anti-epilepsi, macrocrystallin, bio-equival, theopyllin	&	macrocryst, anti-epilepsi, macrocrystallin, \colorbox{cyan}{brand-nam}, bio-eq	\\
117	&	capac, u.s., cellular, telephon, network	&	call-handl, vehicle-bas, areturn, phone, microcel	&	phone, vehicle-bas, call-handl, areturn, microcel	\\
118	&	intern, terrorist	&	terror, osumu, terrorist-track, iraq-inspir, nidal	&	terror, nidal, terrorist-track, \colorbox{cyan}{lebanon-bas}, \colorbox{cyan}{bouzildi}	\\
119	&	action, against, intern, terrorist	&	terror, lybyan, respons, retali, contingency-measur	&	terror, nidal, retali, \colorbox{cyan}{libya}, \colorbox{cyan}{bomb}	\\
120	&	econom, impact, intern, terror	&	consequ, 90-nation, moreov, wage-set, economi	&	consequ, humanwav, \colorbox{cyan}{terrorist}, inde, \colorbox{cyan}{iraq-inspir}	\\ \hline
\multicolumn{4}{|c|}{FastText} \\ \hline
qid	&	query	&	AWE (AQE-Cent)	&	IDF-AWE	(IDF-AWE-VS + AQE-IDF-Cent$^+$) \\ \hline
101	&	design, star, war, anti-missil, defens, system	&	bookstor, armament, staunch, singer, troop	&	antimissil, anti-lock, warhead, \colorbox{cyan}{anti-submarin}, \colorbox{cyan}{anti-tank}	\\
102	&	laser, research, applic, u.s., strateg, defens	&	us, military-tech, research-and-dev, manufacturing, businessmen	&	\colorbox{cyan}{computer-design}, lasertech, \colorbox{cyan}{prospect}, technic, practic	\\
103	&	welfar, reform	&	welfare-reform, welfare-to-work, price-reform, bureau, work-for	&	welfare-reform, welfare-to-work, work-for-welfar, bureau, \colorbox{cyan}{reagan}	\\
104	&	catastroph, health, insur	&	health-insur, health-car, health-polici, catastrophic-il, disease-associ	&	\colorbox{cyan}{fatal}, disease-associ, accid, diseas, \colorbox{cyan}{stroke}	\\
105	&	black, monday	&	black-and-whit, black-and-blu, sunday, friday, bush	&	black-and-whit, gray, sunday, black-and-blu, friday	\\
106	&	u.s., control, insid, trade	&	offic, government-procur, french, prosecut, shipment	&	\colorbox{cyan}{insider-trad}, offic, \colorbox{cyan}{export-control}, incent, outsid	\\
107	&	japanes, regul, insid, trade	&	trade-regulatori, regulatori, american-control, trading-surveil, polit	&	grensid, insider-trad, regulatori, offic, trade-regulatori	\\
108	&	japanes, protectionist, measur	&	protest, japanese-own, american-design, internation, american	&	protest, \colorbox{cyan}{internation}, american-design, \colorbox{cyan}{assault}, assail	\\
109	&	find, innov, compani	&	discoveri, procur, u.s.-develop, develop, disclos	&	non-innov, invis, indirect, discoveri, \colorbox{cyan}{emerg}	\\
110	&	black, resist, against, south, african, govern	&	southern, west, eastern, north, oppress	&	southern, west, eastern, north, northern	\\
111	&	nuclear, prolifer	&	non-nuclear, protein, cancer, geneticist, cooper	&	cancer, protein, geneticist, \colorbox{cyan}{cancer-rel}, regulatori	\\
112	&	fund, biotechnolog	&	project, swedish, nation, research, ventur	&	bio-technolog, technolog, \colorbox{cyan}{biopharmaceut}, biolog, \colorbox{cyan}{pharmaceut}	\\
113	&	new, space, satellit, applic	&	promis, design, sophist, continu, real	&	\colorbox{cyan}{military-satellit}, \colorbox{cyan}{environmental-satellit}, \colorbox{cyan}{spy-satellit}, space-st, sat	\\
114	&	non-commerci, satellit, launch	&	satellite-launch, relaunch, military-satellit, launcher, satellit	&	commerci, commercial-satellit, commerc, \colorbox{cyan}{satellite-launch}, launcher	\\
115	&	impact, 1986, immigr, law	&	lawyer, heard, 1979, ferguson, 1948	&	\colorbox{cyan}{emigr}, \colorbox{cyan}{ancestri}, countri, confus, democraci	\\
116	&	generic, drug, substitut	&	drug-us, prescription-drug, therapeut, drugstor, medication-substitut	&	medication-substitut, \colorbox{cyan}{generic-drug}, prescription-drug, latest, antibiot	\\
117	&	capac, u.s., cellular, telephon, network	&	cellular-servic, portug, consortium, servic, offic	&	cellular-servic, metrophon, creation, portug, so-cal	\\
118	&	intern, terrorist	&	terrorism-watch, surrend, contemporari, conting, indign	&	terrorism-watch, \colorbox{cyan}{terrorist-track}, \colorbox{cyan}{anti-terrorist}, tourist, terror	\\
119	&	action, against, intern, terrorist	&	internecin, sanction, intervent, fervent, incent	&	terrorism-watch, terrorist-track, \colorbox{cyan}{anti-terrorist}, sanction, vehement	\\
120	&	econom, impact, intern, terror	&	econometr, economist, incent, extern, profit	&	incent, \colorbox{cyan}{profit}, economist, poverti, surrend	\\ \hline
    \end{tabular}
    \end{adjustbox}}
    \label{tab:3}
\end{table}

Table \ref{tab:3} shows some highlighted words that we found interesting. These words have the particularity of identifying new terms related to the original query that expands its meaning. Some words are more specific, while others incorporate related senses. The ability of these embeddings to identify collocations strongly related to the query, such as nuclear-weapon or computer-design, is illustrated. The presence of these words is more significant in FastText than in Skip-grams (see, for example, insider-trad, environmental satellite, anti-submarine, anti-tank, among others). It is also observed that AQE-Cent identifies more specific words when using Skip-grams, which is an effect attributable to the training of this strategy in the local corpus of TREC. Although FastText is trained in an external corpus, it manages to identify several words relevant to the original queries. Its ability to generalize in conjunction with its coding based on sub-words helps the AQE strategy.
\vspace{-3mm}

\section{Conclusion}

We have introduced IDF-AWE, a query vector representation that is useful for AQE. Its effectiveness, in conjunction with FastText, shows that word embeddings are useful in AQE.

We are expanding our work to study its effectiveness using other word embeddings, such as BERT \citep{DevlinCLT19} or relevance-based word embeddings \citep{ZamaniC17}. An alternative of particular interest is to learn a combination of weights that generates a better representation of the query. An approach based on machine learning could be beneficial in this line of research.

\section{Acknowledgements}

Mr. Silva and Dr. Mendoza acknowledge funding support from the Millennium Institute for Foundational Research on Data. Dr Mendoza was funded by ANID PIA/APOYO AFB180002 and from ANID FONDECYT grant 1200211.

\bibliographystyle{unsrtnat}
\bibliography{template}

\end{document}